# Reconstruction of gene regulatory network of colon cancer using information theoretic approach

Khalid Raza[#1], Rafat Parveen[*2]

[#]*Department of Computer Science*
*Jamia Millia Islamia (Central University), New Delhi-110025, India.*
[1]kraza@jmi.ac.in

[*] *Department of Information Systems*
*King Abdulaziz University, KSA*
[2]rafatparveen@yahoo.co.in

***Abstract*— Reconstruction of gene regulatory networks or 'reverse-engineering' is a process of identifying gene interaction networks from experimental microarray gene expression profile through computation techniques. In this paper, we tried to reconstruct cancer-specific gene regulatory network using information theoretic approach - mutual information. The considered microarray data consists of large number of genes with 20 samples - 12 samples from colon cancer patient and 8 from normal cell. The data has been preprocessed and normalized. A t-test statistics has been applied to filter differentially expressed genes. The interaction between filtered genes has been computed using mutual information and ten different networks has been constructed with varying number of interactions ranging from 30 to 500. We performed the topological analysis of the reconstructed network, revealing a large number of interactions in colon cancer. Finally, validation of the inferred results has been done with available biological databases and literature.**

## I. Introduction

Cancer, medically known as a 'malignant', is a kind of disease involving unregulated cell growth. The cancerous cells are divided and grown uncontrollably forming malignant tumors and infest the nearby part of the body. The possible means to diagnose cancer are chemotherapy, radiotherapy and surgery but unfortunately, these methods of treatment often damage healthy cells and tissues. Therefore, identification of molecular markers of cancers may be an alternative approach to diagnose the human cancer and might be useful for development of novel therapies. Although, various significant genes responsible for the genesis of different tumors have been revealed but fundamental molecular interactions are still unclear and remains a challenge for the researchers.

Due to rapid growth in microarray technology, gene expression of tens of thousands of genes can be measured simultaneously in a single experiment using a small amount of test sample that enable the researchers detect cancerous molecular markers [1]. Microarrays have been successfully used in many biomedical applications such as gene discovery, disease diagnosis, drug discovery and toxicology. A typical microarray gene expression data is a matrix $R$ with $N$ rows and $M$ columns, where rows represent genes and column as samples (or environmental conditions or time-point). Due to experimental limitations, major problem with microarray data are dimensionality problem (M<<N) and presence of noise in expression values.

Microarray-based cancer prediction is new and growing area of research. A gene regulatory network (GRN) tries to model the complex regulatory interactions within the living cells and give a realistic representation of gene regulation. The inference of GRN from microarray is referred as 'reverse-engineering. Microarray gene expression profiles of whole genome can be used to understand cancer and to reconstruct cancer-specific GRN. The changes in expression profile of genes across various samples provide information that can be used to filter differentially expressed genes between normal and tumor samples and helps to find regulatory relationships between gene-pairs which lead to the reconstruction of GRN. Mapping the topology of GRNs is a central issue in systems biology research [2]. Also, accurate computational methods to reconstruct genome-scale GRN from gene expression profiles are required to explore these experimental data in new and more integrative way.

Many computational methods have been proposed in the literature to model GRNs including directed graph, Boolean networks, generalized Bayesian networks, linear and non-linear ordinary differential equations (ODEs), machine learning approach, and so on. An extensive review can be found in [2, 3, 5, 6, 15]. Many others have tried to reconstruct cancer-specific GRNs using gene expression profiles [4][7]. In [4], a cancer-specific (prostate cancer) GRN has been reconstructed using Pearson's correlation coefficient (PCC) and a network of few most significant genes and their interactions has been identified. A comprehensive comparative evaluation of many state-of-the-art GRN inference methods has been done by Madhamshettiwar et. al. [7]. Finally best-performing method has been applied to infer GRN of ovarian cancer. Many other attempts has been made to reconstruct GRN of various cancers including colon cancer, ovarian cancer, lungs cancer and breast cancer.

In this work, information theoretic approach called mutual information has been used to compute regulatory relationships between gene-pairs. We applied this approach to reconstruct

GRN of colorectal cancer (CRC), the third leading cause of cancer mortality world-wide, which is a genetic disease, propagate by the acquisition of somantic alternations that influence the expression level of gene.

## II. Materials and Methods

In systems biology, many gene regulatory network (GRN) inference methods use information theoretic approach as an estimator to unveil the interaction and relations among genes in a cellular system from gene expression profiles. One of the initial method based on mutual information for GRN inference was introduced in [8]. This method, called relevance network (RN), assigns edges to gene pairs if the corresponding MI value is above a given threshold. The networks that result from application of RN are association networks because an edge between two genes indicates their association but not necessarily a causal effect. Another types of inference methods are also available that intend to find out causal interactions among genes and their products which can be validated with biological experiments, available databases and literatures. Till now, there is no generally agreed gold standard to conduct and include the routine of gene regulatory network inference and their analysis for molecular studies. However, essential preprocessing steps of the data are required to prepare them for the subsequent inference of a gene regulatory network involve standardized procedures for the normalization of gene expression distributions within and between samples and a summarization step to obtain gene-centric values of the gene expression [9].

In this paper, we have applied mutual information to find the regulatory relationship between gene-pairs.

### *A. Algorithm for the reconstruction of gene regulatory network*

The main steps for the reconstruction of gene regulatory networks are outlined as follows:

(1) Data preprocessing and normalization.
(2) Identification of significant genes.
(2) Computation of MI among gene pairs.
(3) Elimination of week correlation links.
(4) Computation of adjacency matrix and network generation.
(5) Biological validation of the results.
(6) Druggability analysis

An sketch of the proposed method is shown in Fig. 1.

### *B. Identification of significant genes*

Before gene expression data is analyzed, first it is ensured data the data set includes genes that differ in their expression level significantly between two classes of samples. Many methods are available for identification of differentially expressed genes in the literature including fold-change, t-test statistics, ANOVA [10] rank product [11], Significant Analysis of Microarray (SAM) [14], Random Variance Model (RVM) [12], Limma [1], and so on. Due to wide applications and significant results of t-test statistics for samples having two different classes (e.g. cell types, cancer types, experimental conditions), we applied t-test to identify differentially expressed between the two classes normal and tumour. The t-test for unpaired data and both for equal and unequal variance can be computed as [4],

$$t_i = \frac{\bar{y}_i - \bar{x}_i}{\sqrt{\frac{g_i^2}{n_1} + \frac{h_i^2}{n_2}}} \qquad (1)$$

where $x_i$ and $y_i$ are the means, $g_i$ and $h_i$ are the variances, and $n_1$ and $n_2$ are the sizes of the two groups of the sample (conditions) tumor and normal, respectively, of gene expression profile $i$.

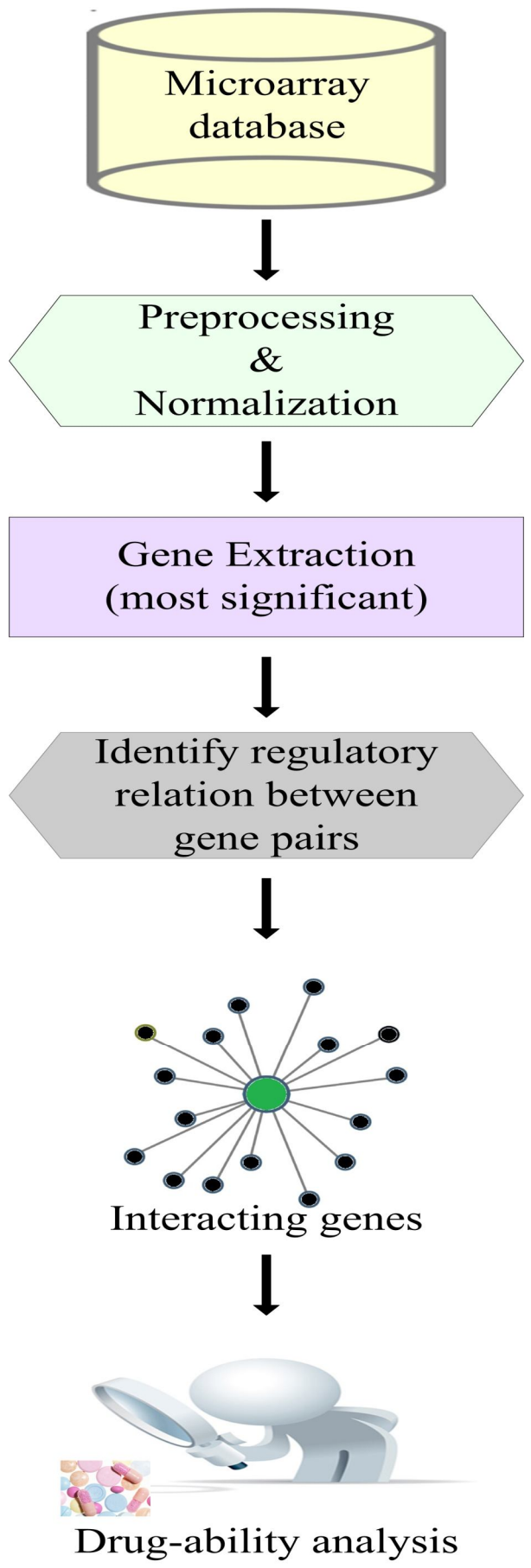

Fig. 1 Steps of the proposed methodology

### *C. Estimation of Mutual Information*

The mutual information (MI), based on information theory, is a general measure for the nonlinear dependence of the two random variables. It is generalisation of pairwise correlation coefficient used to compare expression profiles of a set of microarrays and to measure the degree of independence between two genes. For each pair of genes, their MI($x$,$y$) is computed and the edge $a_{xy}$=$a_{yx}$ is set to 0 or 1 depending on a significant threshold. Mutual information, MI($x$,$y$), between gene $x$ and gene $y$ is computed as:

$$MI(x,y) = H(x) + H(y) - H(x,y) \qquad (2)$$

where entropy $H$ can be defined as:

$$H(x) = -\sum_{k=1}^{n} p((x_k)\log(p(x_k)) \qquad (3)$$

and the joint entropy $H(x,y)$ is defined as,

$$H(x,y) = -\sum_{x_i \in X}\sum_{y_j \in Y} p(x_i, y_j)\log(p(x_i, y_j)) \quad (4)$$

From the definition, the MI becomes zero if the two random variables $x$ and $y$ are statistically independent [that is, $p(x,y)=p(x)p(y)$], as their joint entropy $H(x,y)= H(x)+H(y)$. A higher MI value specifies that the two genes are non-randomly connected to each other. The MI describes an undirected graph because it is symmetric, $MI(x,y)=\mathrm{MI}(y,x)$. MI is more generalized than the Pearson correlation coefficient (PCC) because it quantifies only linear dependencies between variables. However, MI and PCC yield almost identical results. According to the definition of MI, it requires each samples (experiment) to be statistically independent from the others and thus this approach can deal with steady-state as well as with time-series gene expression data.

## III. Results and Discussions

In the present study we took the microarray data of circulating plasma RNA dataset of colorectal cancer (CRC) patient consisting of 20 samples collected from CRC patients. Out of 20 samples, 12 are from colon tumors and 8 are from normal biopsies. The dataset contains the expression profiles of 15552 genes obtained by measuring the relative abundance of the different RNA species in plasma through cDNA microarray hybridization, by comparing RNA isolation and amplified from colorectal cancer patients and from healthy donors. We downloaded the full data set from Gene Expression Omnibus (GEO) [13]. Fig. 2 shows the comparative view of gene expression of different samples for both colon cancer samples and normal samples. The expression of sample profiles in normal tissue is higher in comparison to that of cancer tissue in most of the cases. Many of the cancer sample profiles are down-regulated, as shown in Fig. 2.

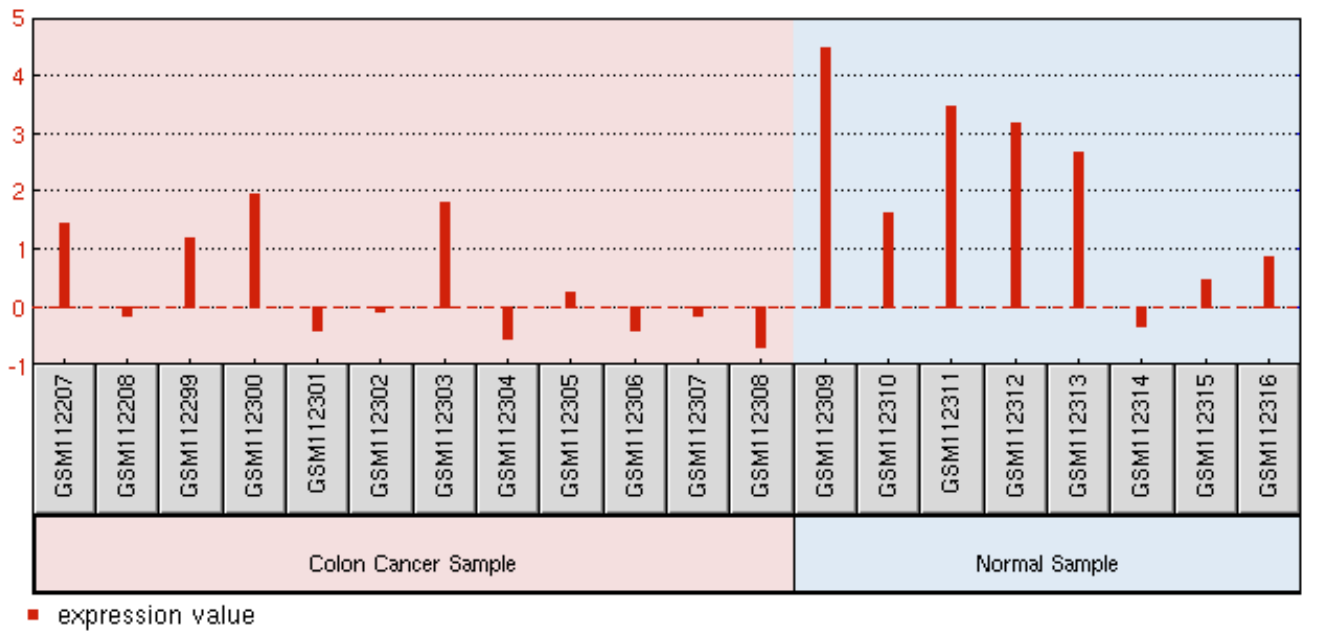

Fig. 2 Sample profile graph showing expression values in cancer and normal sample.

Gene expression data contains a large number of genes, the majority of which may not be relevant for analysis. We applied t-test statistics to select most significant genes from the above dataset whose p-values are less than 0.01. We also eliminated those genes whose either gene name is not available or most of the values in expression profiles are missing. In this way, we found 101 most significant genes that have been selected for further analysis. To find the regulatory interactions among the selected significant genes, mutual information between gene pairs has been computed using equation (2). Now gene interaction network has been constructed, where nodes correspond to gene names and pair-wise mutual information is allocated to the edge between genes. Initially, we took the top 30 highest pair-wise MI values (can be assumed as interaction weight) for the network construction and found a network of 22 genes, which is shown in Fig. 3. From the Fig. 3, it is clear that gene ACAT2 is highly connected with a degree of 9 and regulating a large number of genes. Similarly, we constructed nine other network by taking top 40, 50, 60, 70, 80, 90, 100, 250 and 500 MI values and observed the five highly connected genes in each case. The observation of five highly connected genes in each of the network is shown in Table 1. From the Table 1, it is clear that as the number of interactions are increases, the degree of each hub genes are increases. The network 10 considers 500 interactions that involves 79 genes, in which five highly connected genes are ACAT2(54), CYP1B1(50), NPM1(48), COX15(46), CREM(42), where numbers in parenthesis shows the connection degrees. The network 10 is shown in Fig. 4.

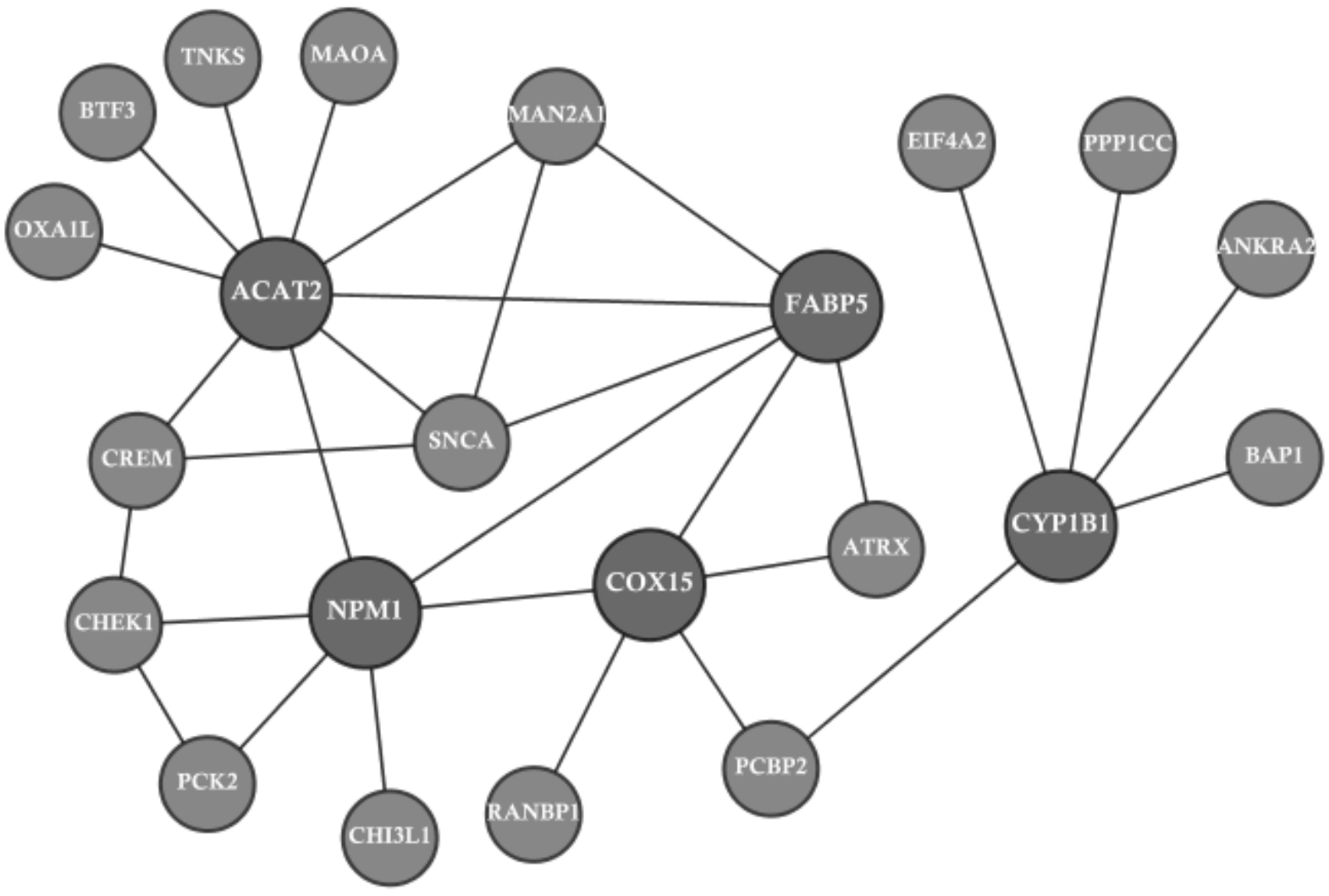

Fig. 3 Network of 22 genes and 30 interactions

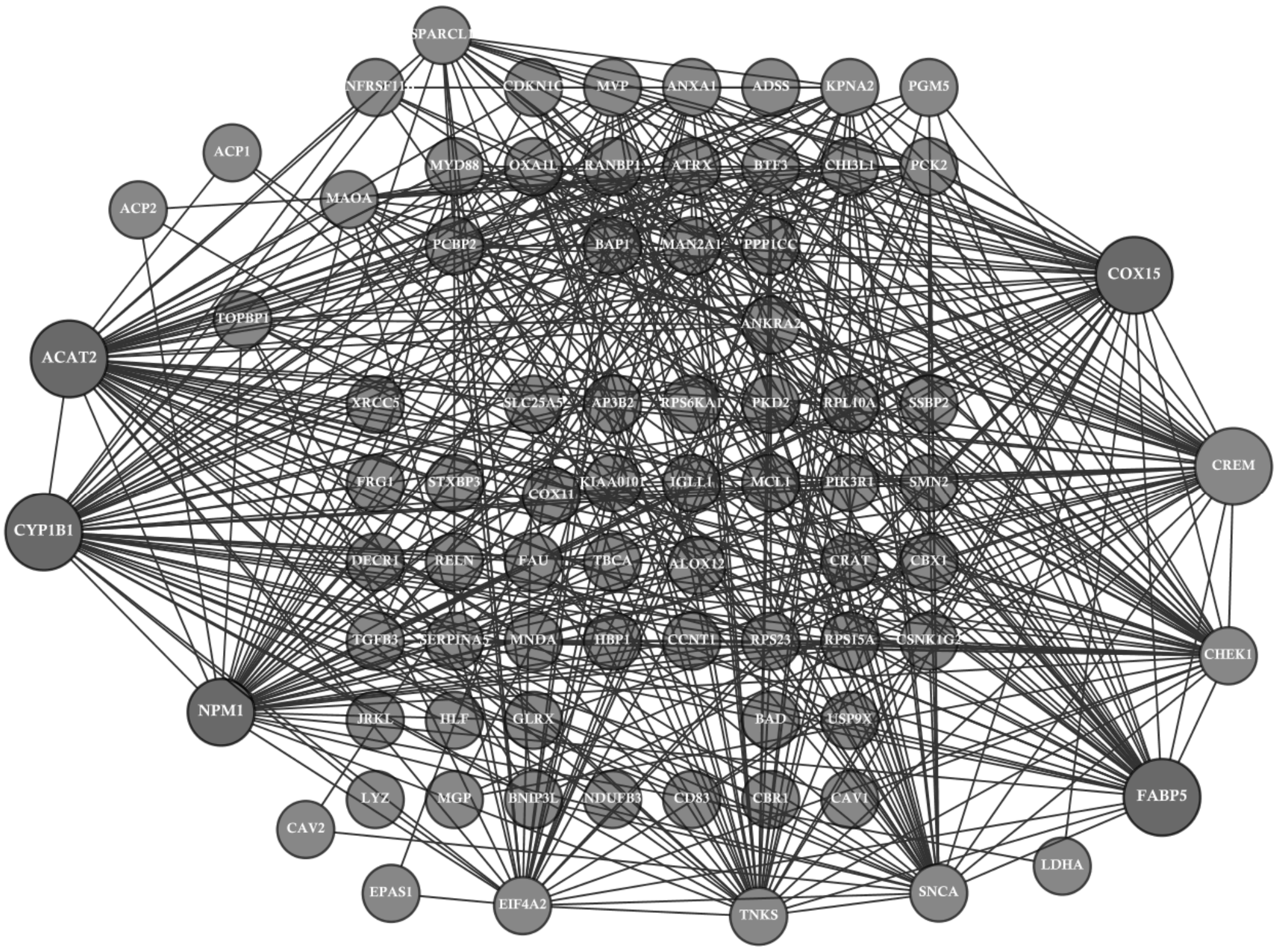

Fig. 4 Network of 79 genes and 500 interactions

Table 1. Ten different networks, number of genes involved in each, five highly connected genes with their degrees.

| Network No. | No. of interactions | Number of genes | Top five genes with highest degree |
|---|---|---|---|
| Network 1 | 30 | 22 | ACAT2 (9), FABP5 (6), NPM1(6), COX15(5), CYP1B1(5) |
| Network 2 | 40 | 25 | ACAT2(9), FABP5(9), NPM1(8), CYP1B1(8), CREM(6) |
| Network 3 | 50 | 27 | ACAT2(12), NPM1(11), CYP1B1(10), FABP5(10), SNCA (7) |
| Network 4 | 60 | 29 | ACAT2(14), NPM1(13), CYP1B1(11), FABP5(10), SNCA(7) |
| Network 5 | 70 | 30 | ACAT2(14), NPM1(13), COX15(12), CYP1B1(12), FABP5(10) |
| Network 6 | 80 | 31 | CYP1B1(16), ACAT2(14), NPM1(13), COX15(12), SNCA(11) |
| Network 7 | 90 | 35 | ACAT2(17), CYP1B1(16), NPM1(14), SNCA(13), TNKS(13) |
| Network 8 | 100 | 36 | NPM1(18), ACAT2(17), CYP1B1(16), SNCA(13), TNKS(13) |
| Network 9 | 250 | 56 | CYP1B1(34), NPM1(33), COX15(30), ACAT2(30), CREM(28) |
| Network 10 | 500 | 79 | ACAT2(54), CYP1B1(50), NPM1(48), COX15(46), CREM(42) |

The identification of highly connected genes (hubs) may play a vital role in cancer diagnosis and therapies. The extracted genes has been validated with the available biological databases and literatures and found that most of the identified genes including ACP, LDHA, SPARCL, EPAS1, MVP, OXA1L, RPL10A, etc. are involved in colon cancer. All the identified interaction among genes including highly connected genes needs biological validation for its reliability. Further, the proposed method can be applied to benchmark as well simulated dataset for its accuracy measure.

## IV. Conclusions

Our study shows application of information theoretic approach to colon cancer, demonstrating how this approach can reveal novel gene regulatory interactions in case of cancer. We constructed ten different networks by varying the number of interactions ranging from 30 to 500, as shown in Table 1. The identified signature in first network captures the regulatory relationships among 22 differentially expressed genes. In case of tenth network considering 500 interactions, it shows regulatory relationships among 79 differentially expressed genes. Our study resulted three major outcomes. First, we identified differentially expressed genes in colon cancer patient, most of them are biological verified and found to participate in colon cancer. Second, the interactions between differentially expressed genes has been identified, which needs further biological validation. Third, we identified genes regulating most of the other genes (hubs). The utility of our approach and the reliability of the obtained results needs further experimental validation. These findings may help to reveal the common interaction mechanism of colon cancer and provide new insights into cancer diagnostic and therapy.